\documentclass[aps, preprint, groupedaddress, floatfix, longbibliography]{revtex4-2}
\usepackage{titlesec}
\usepackage{epsfig}
\usepackage{epstopdf}
\usepackage{graphicx}
\usepackage{amsmath}   
\usepackage{amsfonts}   
\usepackage{color}
\usepackage[export]{adjustbox}
\usepackage[colorlinks=true, allcolors=blue]{hyperref}
\usepackage{lineno}
\begin{document}

\title{Origin of local magnetic exchange interaction in infinite-layer nickelates}

\author{Yanbing Zhou$^{1}$\footnote{These authors contributed equally to this work.}, Dan Zhao$^{1}$\footnotemark[1], Boyun Zeng$^2$\footnotemark[1], Chengliang Xia$^{2}$, Yu Wang$^{1}$, Hanghui Chen$^{2,3}$\footnote{Correspondence to: hanghui.chen@nyu.edu, wutao@ustc.edu.cn, chenxh@ustc.edu.cn}, Tao Wu$^{1,4,5,6}$\footnotemark[2], Xianhui Chen$^{1,4,5,6}$\footnotemark[2]}
\affiliation{$^1$Hefei National Research Center for Physical Sciences at the Microscale, University of Science and Technology of China, Hefei 230026, China\\
	$^2$NYU-ECNU Institute of Physics, NYU Shanghai, Shanghai 200124, China\\
	$^3$Department of Physics, New York University, New York, New York 10012, USA\\
	$^4$Department of Physics, University of Science and Technology of China, Hefei 230026, China\\
	$^5$Collaborative Innovation Center of Advanced Microstructures, Nanjing University, Nanjing 210093, China\\
	$^6$Hefei National Laboratory, University of Science and Technology of China, Hefei 230088, China}

\begin{abstract}
	\textbf{Significant magnetic exchange interactions have been observed in
	infinite-layer nickelates $R$NiO$_2$ ($R$ = La, Pr, Nd), which
	exhibit unconventional superconductivity upon hole doping. Despite
	their structural and Fermi surface similarities to cuprates,
	infinite-layer nickelates possess a larger charge transfer gap, which
	influences their magnetic exchange interactions via oxygen.  In this
	work, we performed $^{17}$O nuclear magnetic resonance (NMR)
	measurements on LaNiO$_2$ and
	Sr-doped LaNiO$_2$, revealing glassy spin dynamics originating from Ni-O planes. This indicates that infinite-layer nickelates are in
	proximity to magnetic ordering and that magnetic correlations play a
	crucial role in their physics.  More importantly, our analysis of the
	Knight shift and hyperfine coupling of $^{17}$O nuclei revealed that the
	Ni-Ni superexchange interaction, mediated by the $\sigma$ bond between the Ni-$d_{x^2-y^2}$ and O-$p$ orbitals, is one order of magnitude weaker than that in
	cuprates. This alone cannot account for the total magnetic exchange
	interaction observed in nickelates. First-principles many-body
	calculations indicate that an interstitial $s$ orbital near the Fermi
	level, coupled with the Ni-$d_{3z^2-r^2}$ orbital, significantly
	enhances the superexchange interaction. This contrasts with cuprates,
	where magnetic interactions are predominantly governed by
	Cu-$d_{x^2-y^2}$ superexchange via oxygen. Our findings provide new
	insights into the distinct magnetic interactions in infinite-layer
	nickelates and their potential role in unconventional
	superconductivity.}
\end{abstract}

\maketitle

\newpage


\section*{Introduction}

The discovery of superconductivity in infinite-layer nickelates $R$NiO$_2$ ($R$ = La, Pr, Nd), has generated significant interest owing to their structural and electronic similarities to cuprates~\cite{li2019superconductivity,osada2021nickelate,zeng2022superconductivity,osada2020superconducting}. However, key differences between these two material families were recognized early in the study of nickel-based superconductors. One major distinction is that infinite-layer nickelates exhibit a substantially larger charge transfer gap than cuprates. As a direct consequence, the magnetic exchange interaction between neighboring Ni-$d_{x^2-y^2}$ spins—mediated via electron exchange through O-$p$ orbitals (known as superexchange)—is expected to be significantly weaker than in cuprates. Reference~\cite{jiang2020critical} explicitly demonstrated that the $\sigma$-bond-driven superexchange in infinite-layer nickelates is an order of magnitude smaller than that in cuprates.

Nevertheless, strong antiferromagnetic correlations have been observed in infinite-layer nickelates via nuclear magnetic resonance (NMR), muon spin rotation ($\mu$SR), magnetic susceptibility and electronic Raman spectroscopy~\cite{zhao2021intrinsic,fu2019core,ortiz2022magnetic,fowlie2022intrinsic,lin2022universal}. More importantly, resonant inelastic X-ray scattering (RIXS) measurements reveal dispersive magnetic excitations with a sizable exchange coupling of $J\sim64$ meV between nearest-neighbor Ni spins—approximately half the value found in cuprates~\cite{lu2021magnetic,krieger2022charge,gao2024magnetic,fan2024capping,hayashida2024investigation}. This apparent discrepancy between the theoretical predictions and experimental observations raises several fundamental questions: i) What role do O-$p$ orbitals play in the superexchange mechanism of infinite-layer nickelates?  ii) Is the total magnetic exchange coupling in nickelates solely governed by the $\sigma$ bond between the Ni-$d_{x^2-y^2}$ and O-$p$ orbitals, or do additional mechanisms contribute?  iii) If other mechanisms are involved, which orbitals participate in the exchange process?  

In this work, we combine NMR measurements with first-principles many-body calculations to address these questions. By performing $^{17}$O NMR measurements on polycrystalline LaNiO$_2$ and La$_{1-x}$Sr$_x$NiO$_2$ samples, we found that the Ni‒Ni superexchange interaction mediated by the $\sigma$ bond between the Ni-$d_{x^2-y^2}$ and O-$p$ orbitals accounts for only approximately 10\% of that in cuprates, which is consistent with the theoretical expectations of Ref.~\cite{jiang2020critical}.
This suggests that the $\sigma$-bond-driven superexchange alone is insufficient to explain the total magnetic exchange coupling observed in infinite-layer nickelates~\cite{lu2021magnetic,krieger2022charge,gao2024magnetic,fan2024capping,hayashida2024investigation}.
Furthermore, our first-principles many-body calculations reveal that the presence of an interstitial $s$ orbital near the Fermi level, which hybridizes with the Ni-$d_{3z^2-r^2}$ orbital, significantly enhances the superexchange in infinite-layer nickelates. This additional contribution to magnetic exchange is comparable in magnitude to the well-established $d_{x^2-y^2}$-$p$ $\sigma$-bond mechanism. Our results underscore the crucial role of low-energy interstitial $s$ orbitals in the magnetic exchange interactions of infinite-layer nickelates, providing new insights into superexchange mechanisms of unconventional oxide superconductors.


\section*{$^{17}$O NMR spectroscopy}

To conduct $^{17}$O NMR measurements, we first synthesized an $^{17}$O-enriched LaNiO$_3$ polycrystalline sample. Then, using CaH$_2$ as a reducing agent, the $^{17}$O-enriched LaNiO$_3$ precursor is converted into the final $^{17}$O-enriched LaNiO$_2$ polycrystalline sample [for details, see Sec. S1 in the Supplemental Information]. As shown in Figure~\ref{fig1}(a), there is only one structural site for oxygen atoms in both stoichiometric LaNiO$_3$ and LaNiO$_2$, which simplifies the present $^{17}$O NMR investigation. A typical high-temperature $^{17}$O NMR spectrum of the LaNiO$_2$ polycrystalline sample is shown in Fig.~\ref{fig1}(c), 
which is significantly different from the $^{17}$O NMR spectrum of polycrystalline LaNiO$_3$ (Fig.~\ref{fig1}(b)). Considering the high spin number $I=5/2$ of $^{17}$O nuclei, both the quadrupole and magnetic hyperfine interactions experienced by $^{17}$O nuclei determine the characteristics of the polycrystalline NMR spectrum. Our simulation of the polycrystalline NMR spectrum provides a quantitative understanding of the $^{17}$O NMR spectra of both LaNiO$_2$ and LaNiO$_3$, which indicates that the NMR parameters (electric field gradient (EFG) tensor and Knight shift) in LaNiO$_2$ are more anisotropic than those in LaNiO$_3$ [for details, see the captions of Fig.~\ref{fig1} and Sec. S2 in the Supplemental Information]. In addition, our simulation also indicates that the distribution of the quadrupole frequency ($\frac{\Delta \nu_{q}}{\nu_{q}}$) increases from $\sim 2$\% to $\sim 6$\% during the change from LaNiO$_3$ to LaNiO$_2$, suggesting increased inhomogeneity from structural or charge degrees of freedom. As the temperature decreases, the $^{17}$O NMR spectra exhibit a significant broadening effect at low temperatures (Fig.~\ref{fig1}(d)), which smears out the characteristics of the $^{17}$O NMR spectra and leads to increased uncertainty in the extracted NMR parameters. Therefore, we directly used the second central moment ($M_2$) of the $^{17}$O NMR spectra to investigate the low-temperature broadening effect. As shown in Fig.~\ref{fig1}(e), the temperature-dependent second central  moment rapidly increases below $\sim 70$ K, suggesting possible electronic inhomogeneity. 
With the help of simulations of the broadening effect (Fig.~\ref{fig1}(f)), we conclude that the origin of such a broadening effect is dominated by magnetic contributions. In contrast, if the quadrupole contribution is predominant for the broadening effect, then the $^{17}$O NMR spectrum always has a sharp central peak (Fig.~\ref{fig1}(f)), which is inconsistent with our observations. Notably, strictly speaking, we cannot exclude a secondary quadrupole contribution to the total broadening effect. Our results clearly show that a predominant magnetic broadening effect appears in the $^{17}$O NMR spectra below $\sim 70$ K, suggesting possible magnetic inhomogeneity at low temperatures. Previous AC susceptibility and $\mu$SR measurements of a LaNiO$_2$ polycrystalline sample revealed spin‒glass behavior at low temperatures~\cite{ortiz2022magnetic,lin2022universal}, which provides a natural explanation for the observed magnetic broadening effect in the $^{17}$O NMR spectra. Furthermore, we confirm that the magnetic broadening effect indeed comes from spontaneous internal fields due to the spin glass state, which leads to almost field-independent $^{17}$O NMR spectra at low temperatures [for details, see Sec. S3 in the Supplemental Information]. In this work, we also synthesized an $^{17}$O-enriched polycrystalline La$_{0.82}$Sr$_{0.18}$NiO$_2$ sample and performed $^{17}$O NMR measurements. In Sr-doped LaNiO$_2$, a similar spin‒glass state is also observed in the temperature-dependent $^{17}$O NMR spectra [for details, see Sec. S4 in the Supplemental Information]. Compared with that of pristine LaNiO$_2$, the high-temperature $^{17}$O NMR spectrum of La$_{0.82}$Sr$_{0.18}$NiO$_2$ has a wider distribution of NMR parameters [for details, see Sec. S4 in the Supplemental Information], suggesting an enhanced disorder effect due to the Sr dopants.

\section*{Glassy spin dynamics}

To further investigate the spin-glass behavior, we measured the temperature-dependent spin-lattice relaxation rate ($1/T_1$), which probes the spectral density of spin fluctuations at the NMR frequency. Figure~\ref{fig2}(a) shows the typical recovery curves of $^{17}$O nuclear magnetization in both LaNiO$_2$ and La$_{0.82}$Sr$_{0.18}$NiO$_2$. To fit the recovery curves well, we use a stretched exponential formula, $M(t)=M_0+M_1\left( 0.028e^{-\left(\frac{t}{T_1} \right)^\alpha }+0.178e^{-\left(\frac{6t}{T_1} \right)^\alpha }+0.794e^{-\left(\frac{15t}{T_1} \right)^\alpha } \right)$, which is widely used to describe the glassy spin dynamics of cuprates~\cite{julien1999charge,julien2001glassy,wu2013magnetic,frachet2020hidden,PhysRevB.86.220504,PhysRevLett.85.642,arsenault2018139,PhysRevB.61.R9265,PhysRevB.78.014504,PhysRevB.63.020507,PhysRevB.64.134525,PhysRevB.96.094519,PhysRevB.92.155144,PhysRevB.46.3179,julien2000nqr}. Here, the stretching exponent ($\alpha$) reflects the distribution of spin‒lattice relaxation times ($T_1$) and serves as a key indicator of glassy spin dynamics. In cuprates, the glassy spin dynamics are well captured by the Bloembergen‒Purcell‒Pound (BBP) mechanism~\cite{PhysRevB.106.054522,PhysRevX.15.021010,frachet2020hidden,PhysRevB.78.014504}, which assumes that the autocorrelation function of the fluctuating hyperfine field $h(t)$ decays exponentially with time ($t$): $\left\langle h(t)h(0) \right\rangle=\left\langle h^2\right\rangle e^{-t/\tau_c} $~\cite{bloembergen1948relaxation}. Accordingly, the spin-lattice relaxation rate is given by  $\frac{1}{T_1}=\gamma_N^2\left\langle h_\perp^2\right\rangle\frac{2\tau_c}{1+\omega_{N}^2\tau_c^2}$, where $\tau_c$ is the correlation time between local spins and is assumed to decrease with temperature as $\tau_c=\tau_\infty e^{E/k_{B}T}$. Here, the activation energy ($E$) is assumed to be temperature independent, $h_\perp$ is the transverse fluctuating field at the nuclear site, and $\gamma_N$ is the gyromagnetic ratio of the nuclei. Within this framework, as the correlation time increases with cooling, the value of $1/T_1$ reaches a maximum when $\tau_c\sim\omega_{N}$. The temperature of the maximum $1/T_1$ defines the freezing temperature ($T_{1f}$) on the NMR timescale.

As shown in Fig.~\ref{fig2}(b), the temperature-dependent $1/T_1$ exhibits broad peak-like behavior at low temperatures in both LaNiO$_2$ and La$_{0.82}$Sr$_{0.18}$NiO$_2$, which supports the spin freezing of local spins and defines the characteristic freezing temperature $T_{1f}$ ($\sim52$ K for LaNiO$_2$ and $\sim26$ K for La$_{0.82}$Sr$_{0.18}$NiO$_2$). Moreover, the stretching exponent $\alpha$ becomes much smaller than 1 at low temperatures (Fig.~\ref{fig2}(c)), indicating a remarkable inhomogeneity of spin dynamics in the spin-glass state. On the other hand, the overall temperature dependence of $1/T_1$ suggests that the local spins in the system are only partially frozen. By further qualitative analysis [for details, see Sec. S5 in the Supplemental Information], $1/T_1$ can be decomposed into two contributions: one is the spin freezing component ($1/T_1 )_{\mathrm{BPP}}$, which is well described by the BPP model, and the other is a paramagnetic component $(1/T_1 )_{\mathrm{para}}$, which continuously decreases with decreasing temperature and represents unfrozen spins with strong antiferromagnetic correlations~\cite{zhao2021intrinsic}. Notably, while the peak temperature of $1/T_1$ in La$_{0.82}$Sr$_{0.18}$NiO$_2$ is much lower than that in LaNiO$_2$, Sr doping significantly enhances the magnitude of the peak in $1/T_1$ (Fig.~\ref{fig2}(b)), implying an increase of the frozen spins. This observation suggests a dual role of Sr doping in glassy spin dynamics. On the one hand, the Sr dopants enhance chemical/structural disorder, which leads to an increase of frozen spins. On the other hand, Sr doping introduces additional holes, which alter the activation energy for spin glass. More evidence for glassy spin dynamics is also provided by $^{139}$La NMR, where the peak behavior in $1/T_1$ clearly exhibits frequency dependence, as expected by the BPP model [for details, see Sec. S5 in the Supplemental Information].

Further measurement of the nuclear spin‒spin relaxation rate $1/T_2$, which measures slow longitudinal field fluctuations, also supports a spin‒glass state at low temperatures. The nuclear spin‒spin relaxation time $T_2$ was extracted by fitting the spin-echo decay curves via the formula $S(\tau) = S_{0}e^{- \left( \frac{2\tau}{T_{2}} \right)^{\beta}}$ (Fig.~\ref{fig2}(d)), where the factor $\beta<1$ accounts for a distribution of $T_2$ values~\cite{PhysRevB.103.L180506}. As shown in Fig.~\ref{fig2}(e), the temperature-dependent $1/T_2$ develops a pronounced broad peak at low temperatures, which is a typical characteristic of the spin-glass state. Similar to the results of $1/T_1$, Sr doping leads to an enhanced peak in $1/T_2$, supporting an increase of the frozen spins. In addition, the stretching factor $\beta$ clearly decreases around the peak region (Fig.~\ref{fig2}(f)), which is consistent with glassy spin dynamics.

\section*{Superexchange mediated by the $\sigma$ bond}

The observed spin‒glass behavior in infinite-layer nickelates resembles that in underdoped cuprates~\cite{julien1999charge,julien2001glassy,wu2013magnetic,frachet2020hidden,PhysRevB.86.220504,PhysRevLett.85.642,arsenault2018139,PhysRevB.61.R9265,PhysRevB.78.014504,PhysRevB.63.020507,PhysRevB.64.134525,PhysRevB.96.094519,PhysRevB.92.155144,PhysRevB.46.3179,julien2000nqr}, in which carrier doping suppresses long-range magnetic order but a glassy magnetic state persists~\cite{birgeneau2006magnetic,tacon2011intense,dean2013persistence,minola2015collective,PhysRevB.106.054522,PhysRevX.15.021010,frachet2020hidden,PhysRevB.78.014504}. In infinite-layer nickelates, while self-doping disrupts the establishment of long-range magnetic order~\cite{PhysRevB.101.020501}, significant antiferromagnetic correlations persist, which leads to the observed spin‒glass state~\cite{zhao2021intrinsic,fu2019core,ortiz2022magnetic,fowlie2022intrinsic,lin2022universal,lu2021magnetic,krieger2022charge,gao2024magnetic,fan2024capping,hayashida2024investigation}. Usually, the strong antiferromagnetic correlations among Ni spins are ascribed to the superexchange mediated by the O-$p$ orbital. The present $^{17}$O NMR measurement can offer a unique chance to investigate this kind of superexchange. In 3$d$ transition metal compounds, the effective superexchange interaction strength $J$ is related to the charge transfer gap $\Delta_{pd}$ and is expressed as $J = \frac{{4t}_{pd}^{4}}{\Delta_{pd}^{2}}\left(\frac{1}{U_{dd}} + \frac{1}{\Delta_{pd} + \frac{U_{pp}}{2}}\right)$~\cite{zaanen1987electronic}. Previous $^{17}$O NMR experiments on cuprates have established that the hyperfine coupling constant of $^{17}$O directly reflects the hybridization between O-$p$ and Ni-$d$ orbitals, which is influenced by the charge transfer gap $\Delta_{pd}$~\cite{abragam2013electron}. Therefore, analysis of the hyperfine coupling of $^{17}$O can provide a qualitative estimation of the superexchange interactions mediated by the $\sigma$ bond between the Ni-$d_{x^2-y^2}$ and O-$p$ orbitals in our case.

By analysing the high-temperature $^{17}$O NMR spectra of LaNiO$_2$, we can extract the Knight shifts of $^{17}$O along three crystallographic directions. As shown in Figure~\ref{fig3}(d), the temperature-dependent Knight shift decreases continuously with cooling, which is quite consistent with the previous $^{139}$La NMR results~\cite{zhao2021intrinsic}. In principle, the total Knight shift can be expressed as $K_{i} = K_{i}^{\mathrm{orb}} +
A_{i}\chi_{i}$, where $i = a, b$, and $c$, $K_{i}^{\mathrm{orb}}$ is temperature-independent orbital shift, $A_i$ is the hyperfine coupling constant of $^{17}$O, and $\chi_{i}$ is the bulk spin susceptibility. Specifically, without considering spin‒orbit coupling, the hyperfine coupling constant of $^{17}$O consists of two components: $A_{i} = A^{\mathrm{iso}} + A_{i}^{\mathrm{dip}}$. 
The isotropic term $A^{\mathrm{iso}}$ originates from the Fermi contact interaction due to the hybridization between the Ni-$d$ and O-$2s$ orbitals. The anisotropic term $A^{\mathrm{dip}}$ arises from the dipolar interaction, obeying the relation $A_b^{\mathrm{dip}}=-2A_a^{\mathrm{dip}}=-2A_c^{\mathrm{dip}}$. The dipolar contribution is closely related to the $pd$-hybridization and is given by 
$A_b^{\mathrm{dip}}=2f_{p\sigma}\alpha_{p}$, $A_a^{\mathrm{dip}}=A_c^{\mathrm{dip}}=-f_{p\sigma}\alpha_{p}$, where $\alpha_{p}=\frac{4}{5}\hbar\gamma\mu_{B}\left\langle r^{-3} \right\rangle$ represents the dipolar contribution from the 2$p$ orbital, and $f_{p\sigma}$ is the so-called “spin density”, which gives the statistical weight of the hybridized wave function in the ligand orbitals with $f_{p\sigma}$ character~\cite{abragam2013electron}. Typically, $f_{p\sigma} \propto \frac{t_{pd}^{2}}{\Delta_{pd}^{2}}$, where $t_{pd}$ is the hopping integral between O-$p$ orbitals and Ni-$d_{x^2-y^2}$ orbitals and where $\Delta_{pd}$ is the energy difference between these orbitals~\cite{abragam2013electron}. Given that the superexchange interaction between the localized spins on Ni-$d_{x^2-y^2}$ orbitals is mediated by O-2$p_\sigma$ bonds (Fig.~\ref{fig3}(a)), we can estimate the strength of this superexchange interaction by analysing the hyperfine coupling constants of oxygen $A_i$. By comparing the Knight shifts along different crystallographic directions (Fig.~\ref{fig3}(e)), we determined the anisotropy ratio of the hyperfine coupling constant, $\frac{\Delta K_b}{\Delta
	K_c}=\frac{A_b}{A_c}=\frac{A^{\mathrm{iso}}+A^{\mathrm{dip}}_b}{A^{\mathrm{iso}}-\frac{1}{2}A^{\mathrm{dip}}_b}\sim
1.11$, which is significantly smaller than that of $\sim1.6$ in cuprates (Fig.~\ref{fig3}(f)). Given the comparable $t_{pd}$ values for nickelates and cuprates, this smaller anisotropy ratio suggests a larger charge-transfer gap $\Delta_{pd}$ in nickelates than in cuprates.

To analyse the charge-transfer energy $\Delta_{pd}$ more quantitatively, we measured the bulk magnetic susceptibility $\chi$ and evaluated the $K-\chi$ relationship to extract the hyperfine coupling constants $A_i$ [for details, see Secs. S6 and S7 in the Supplemental Information]. Finally, we obtain $A_a\sim16.1~\mathrm{kOe/\mu_B}$,
$A_{b} \sim
18.9~\mathrm{kOe/\mu_{B}}$, and
$A_c\sim15.2~\mathrm{kOe/\mu_B}$. The dipolar contribution, which is calculated by
$A_b^{\mathrm{dip}}=A_b-\frac{A_a+A_b+A_c}{3}\sim2.2~\mathrm{kOe/\mu_B}$, is much smaller than that in YBa$_2$Cu$_3$O$_\mathrm{7-\delta}$ 
($A_b^{\mathrm{dip}}\sim20.7~\mathrm{kOe/\mu_B}$)~\cite{takigawa1991cu,walstedt2008nmr}. By calculating the ratio $\frac{\Delta_{pd}^{\mathrm{Ni}}}{\Delta_{pd}^{\mathrm{Cu}}}\sim\left(
\frac{(A_b^{\mathrm{dip}})^{\mathrm{Cu}}}{(A_b^{\mathrm{dip}})^{\mathrm{Ni}}}\right)^{\frac{1}{2}}\sim3.1$, we conclude that the charge-transfer gap in LaNiO$_2$ is three times larger than that in cuprates. Furthermore, the ratio of superexchange interactions in cuprates and LaNiO$_2$ can be estimated by $\frac{J^{\mathrm{Cu}}}{J^{\mathrm{Ni}}}=\frac{\left(\frac{1}{\Delta_{pd}^2}\right)_\mathrm{Cu}}{\left(\frac{1}{\Delta_{pd}^2}\right)_\mathrm{Ni}}\cdot\frac{\left(\frac{1}{U_{dd}}+\frac{1}{\Delta_{pd}+\frac{U_{pp}}{2}}\right)_{\mathrm{Cu}}}{\left(\frac{1}{U_{dd}}+\frac{1}{\Delta_{pd}+\frac{U_{pp}}{2}}\right)_{\mathrm{Ni}} }\sim10$. Here, we assume that $U_{dd}^{\mathrm{Ni}}\sim U_{dd}^\mathrm{Cu}$ and that $U_{pp}^{\mathrm{Ni}}\sim U_{pp}^\mathrm{Cu}$. Since the superexchange interaction in cuprates typically ranges from 100 meV to 180 meV (e.g., $J\sim140$ meV in YBa$_2$Cu$_3$O$_\mathrm{7-\delta}$)~\cite{wang2022paramagnons,tacon2013dispersive}, we estimate that the superexchange interaction in LaNiO$_2$ is approximately 10–18 meV. This value is substantially lower than the $J\sim64$ meV extracted from RIXS experiments in both infinite-layer nickelate thin films and bulk samples~\cite{lu2021magnetic,krieger2022charge,gao2024magnetic,fan2024capping,hayashida2024investigation}.

In infinite-layer nickelates, as illustrated in Fig.~\ref{fig3}(b-c), the O-$p$ orbitals are farther from the Fermi level compared to cuprates, and the charge transfer gap $\Delta_{pd}$ is larger than the onsite Coulomb repulsion $U$ of Ni-$d$ electrons~\cite{jiang2020critical,hepting2020electronic,zhao2020electronic}. This difference leads to a larger charge transfer gap $\Delta_{pd}$ in infinite-layer nickelates than in cuprates~\cite{jiang2020critical,hepting2020electronic}. According to the Zaanen–Sawatzky–Allen (ZSA) classification~\cite{zaanen1987electronic}, infinite-layer nickelates are categorized as Mott–Hubbard-type insulators, whereas cuprates are charge-transfer insulators. As $\Delta_{pd}$ increases, it is reasonable to infer that the $\sigma$ bond between the Ni-$d_{x^2-y^2}$ and O-$p_x$ orbitals plays a less significant role in mediating the magnetic exchange interactions in infinite-layer nickelates than does its dominant role in cuprates. Our present $^{17}$O NMR results strongly support this conjecture. Considering that the total superexchange interaction in infinite-layer nickelates has a moderate value ($J\sim64$ meV), we propose that, in addition to the well-established $\sigma$ bond between the Ni-$d_{x^2-y^2}$ and O-$p$ orbitals, other virtual hopping pathways also contribute substantially to the total magnetic interaction in infinite-layer nickelates.

\section*{Theoretical modeling of the superexchange}

We now explain how additional orbitals and extended hopping pathways may appreciably increase the total magnetic exchange of infinite-layer nickelates. Figure~\ref{fig4}(a) shows the well-known $d_{x^2-y^2}$-$p_x$-$d_{x^2-y^2}$ cluster model (hereafter referred to as the $pd$ model) that is widely used to estimate the superexchange mediated by the $pd$ $\sigma$ bond~\cite{zaanen1987electronic}. To better account for the full superexchange in infinite-layer nickelates, we extend the $pd$ model to a new cluster model, shown in Fig.~\ref{fig4}(b). We refer to it as the $pds$ model. Our previous first-principles calculations revealed a key distinction between infinite-layer nickelates and cuprates: the presence of an interstitial $s$ orbital in the former, with its energy lying close to the Fermi level~\cite{gu2020substantial}. In the $pds$ model, we incorporate the interstitial $s$ orbital. In addition, we include the Ni-$d_{3z^2-r^2}$ orbital, motivated by the fact that in the presence of the interstitial $s$ orbital, the bare energies of the $d_{x^2-y^2}$ and $d_{3z^2-r^2}$ orbitals become nearly degenerate~\cite{gu2020substantial,chen2023electronic,PhysRevB.106.035111}. As a result, a new $\sigma$ bonding pathway emerges, formed between the Ni-$d_{3z^2-r^2}$ and interstitial $s$ orbitals, in addition to the well-known $d_{x^2-y^2}$–$p_x$ $\sigma$ bond. We adopt open boundary conditions~\cite{chen2024electronic,PhysRevB.106.115150}, such that each Ni site is flanked by two interstitial $s$ orbitals. All symmetry-allowed nearest-neighbor hoppings are included in the $pds$ model, as explicitly shown in Fig.~\ref{fig4}(b). The onsite orbital energies and hopping parameters are extracted from first-principles calculations and Wannier fitting. We apply a full Slater-Kanamori interaction to each Ni site~\cite{Kanamori1963}. Based on our DMFT calculations, we choose a representative interaction strength of $U = 5.0$ eV and Hund's coupling $J_H = 1.0$ eV, which reproduce the mass enhancement of Ni $d$ orbitals obtained from GW+EDMFT simulations~\cite{PhysRevX.10.041047}. We also explore a range of $U$ and $J_H$ values around these benchmarks. The complete Hamiltonian for the $pds$ cluster model is presented in the Methods. Following Ref.~\cite{zaanen1987electronic}, we compare a ``ferromagnetic'' (FM) configuration with total spin projection $S_z = 1$ and an ``antiferromagnetic'' (AFM) configuration with $S_z = 0$. To represent the formal valence states of Ni$^{1+}$ and O$^{2-}$ in infinite-layer nickelates, we set the total number of electrons in the $pds$ model to $N_e = 8$.

We next perform exact diagonalization of the ``FM'' and ``AFM'' Hamiltonians in the corresponding many-body basis to obtain the respective ground-state energies, \( E_{\mathrm{FM}} \) and \( E_{\mathrm{AFM}} \). Since the model effectively describes two coupled spin-\(\tfrac{1}{2}\) degrees of freedom, the magnetic exchange interaction can be mapped onto a spin-\(\tfrac{1}{2}\) Heisenberg model and the superexchange coupling is obtaine by \( J = |E_{\mathrm{FM}} - E_{\mathrm{AFM}}| \)~\cite{zaanen1987electronic} (see the Methods for more details). Owing to the larger phase space available in the ``AFM'' configuration, we consistently find \( E_{\mathrm{AFM}} < E_{\mathrm{FM}} \). A key result of our analysis is that the energy of the interstitial \( s \) orbital serves as a critical control parameter for tuning the superexchange strength. We define \( \Delta_s \) as the onsite energy difference between the interstitial \( s \) orbital and the Ni-\( d_{x^2 - y^2} \) orbital. In Fig.~\ref{fig4}(c), we plot the dependence of \( J \) on \( \Delta_s \) for several values of the Hubbard interaction \( U \), keeping the Hund's coupling fixed at \( J_H = 0.2U \). A complementary analysis in Fig.~\ref{fig4}(d) explores the variation of \( J \) with \( J_H \) at fixed \( U = 5\,\mathrm{eV} \). In both cases, we find that \( J \) increases significantly as \( \Delta_s \) decreases. In particular, when \( \Delta_s \) is appreciably smaller than \( U \) (shaded region), the interstitial \( s \) orbital becomes energetically relevant and substantially enhances the superexchange. For large \( \Delta_s \) (e.g., \( 10\,\mathrm{eV} \)), the \( s \) orbital lies far above the Fermi level and its contribution to the superexchange vanishes, resulting in a saturated \( J \approx 20\,\mathrm{meV} \). In contrast, when \( \Delta_s \) is set to its first-principles value of approximately \( 2.6\,\mathrm{eV} \), the superexchange increases markedly, with \( J \) reaching values in the range of \( 50\text{--}120\,\mathrm{meV} \), depending on the interaction parameters. While the \( pds \) model is still simplified and not intended to yield quantitative estimate of superexchange, it provides compelling evidence that the combined involvement of the interstitial \( s \) orbital and the Ni \( d_{x^2 - y^2} \) and \( d_{3z^2 - r^2} \) orbitals can strongly enhance the magnetic exchange interactions in infinite-layer nickelates.


\section*{Discussions and conclusions}

In high-$T_c$ superconductors, antiferromagnetic spin fluctuations are widely recognized as central to the mechanism of unconventional pairing~\cite{lee2006doping,RevModPhys.84.1383}. While a precise theoretical prediction of the superconducting transition temperature ($T_c$) remains challenging, numerous studies suggest that the strength of these spin fluctuations imposes an upper bound on $T_c$~\cite{wang2023correlating,wang2022paramagnons}. This has led to the view that enhancing the magnetic exchange interaction $J$---which governs spin fluctuation strength---may be a viable route to higher $T_c$. In cuprates, $J$ arises primarily from superexchange via oxygen $p$ orbitals, and is inversely related to the charge transfer gap $\Delta_{pd}$ between oxygen $p$ and copper $d$ states. The empirical correlation between smaller $\Delta_{pd}$ and higher $T_c$ within the Bi$_2$Sr$_2$Ca$_{n-1}$Cu$_n$O$_{2n+4+\delta}$ series underscores this link~\cite{wang2023correlating}. 

Our study extends this framework to infinite-layer nickelates, where we identify both similarities and crucial differences in the origin of magnetic interactions. In particular, we find that the well-known $d_{x^2-y^2}$-$p$ $\sigma$-bond superexchange is not the dominant contributor to $J$ in these superconductors. Instead, we highlight the significant role of a distinct exchange pathway involving the interstitial-$s$ orbital, which has negligible impact in cuprates due to its high energy. We show that lowering the energy of this orbital---through structural or chemical tuning---can substantially enhance magnetic coupling. This mechanism offers a plausible explanation for the elevated $T_c$ values recently reported in (Sm-Eu-Ca-Sr)NiO$_2$ thin films~\cite{chow2025bulk}, where enhanced interstitial-$s$ participation may amplify the effective superexchange interaction.

In conclusion, our $^{17}$O NMR measurements on infinite-layer nickelates reveal that the conventional in-plane $\sigma$-bond-mediated superexchange is significantly weaker than in cuprates, consistent with the larger charge transfer gap of infinite-layer nickelates. However, this alone cannot account for the full magnitude of the magnetic exchange coupling observed experimentally~\cite{lu2021magnetic,krieger2022charge,gao2024magnetic,fan2024capping,hayashida2024investigation}. Through cluster model calculations, we identify a novel exchange channel involving the interstitial-$s$, Ni-$d_{3z^2 - r^2}$, and O-$p$ orbitals, which contributes substantially to $J$ in nickelates but is negligible in cuprates. These results establish a key distinction in the magnetic physics of nickelates and suggest a broader design principle: magnetic interactions---and by extension, superconductivity---in oxide materials can be engineered by manipulating the energies and hybridizations of non-canonical orbitals near the Fermi level.


\section*{Methods}

\subsubsection*{Sample Preparation}
Powder samples of LaNiO$_3$ and Sr-doped LaNiO$_3$ were synthesized via a sol-gel method~\cite{deganello2009citrate}. Stoichiometric amounts of metal nitrates were mixed to prepare the desired LaNiO$_3$ and La$_{0.8}$Sr$_{0.2}$NiO$_3$ precursor solutions with excess citric acid. The solution was slowly evaporated until it became green gel, which was then burned at $\sim 500^{\circ}$C for several hours to yield gray precursor powder. These precursors of LaNiO$_3$ and La$_{0.8}$Sr$_{0.2}$NiO$_3$ were subsequently calcined at $950^{\circ}$C for 12 h. The LaNiO$_3$ precursor was calcined in flowing O$_2$, whereas the La$_{0.8}$Sr$_{0.2}$NiO$_3$ precursor required an O$_2$ pressure of $\sim$ 120 bar to stabilize the desired perovskite 113 phase. 

For subsequent $^{17}$O NMR measurements, the synthesized oxide powders were $^{17}$O isotopically enriched. Powder LaNiO$_3$ was enriched by annealing at $900^{\circ}$C for 102 h in a $^{17}$O$_2$ atmosphere. Similarly, powder samples of Sr-doped LaNiO$_3$ were enriched by annealing at $710^{\circ}$C for 120 h in a $^{17}$O$_2$ atmosphere. 
Polycrystalline samples of the infinite-layer nickelates, LaNiO$_2$ and Sr-doped LaNiO$_2$, were then obtained by chemical reduction of these $^{17}$O-enriched samples of LaNiO$_3$ and Sr-doped LaNiO$_3$ using CaH$_2$ as the reducing agent~\cite{takamatsu2010low}. The actual stoichiometry of the Sr-doped final product was confirmed to be La$_{0.82}$Sr$_{0.18}$NiO$_2$ through inductively coupled plasma‒mass spectrometry (ICP-MS).

\subsubsection*{NMR Experiments}

To perform NMR measurements on both parent and doped samples, each
isotope enriched sample was filled with a copper coil and then sealed with epoxy adhesive. A commercial NMR spectrometer from Thamway Co. Ltd
was used for the NMR measurements. The external magnetic field at the
sample position was calibrated by the $^{63}$Cu resonance
frequency of the copper coil. All \textsuperscript{139}La and
$^{17}$O NMR signals were measured via the standard spin
echo method using the $\pi/2$-$\tau$-$\pi$-$\tau$ pulse sequence; subsequently, the time-domain signals were converted to spectra via fast Fourier
transform, and these spectra were summed to obtain a full spectrum.

We carried out nuclear spin-lattice relaxation time ($T_{1}$)
measurements on $^{139}$La and $^{17}$O nuclei
using comb $\pi/2$-pulse for saturation, followed by the
$t_j$-$\pi/2$-$\tau$-$\pi$-$\tau$ pulse
sequence, where $t_j$ varies, and we can plot the integral
signal intensity $I_s$ versus $t_j$. Since $I_s$ is
proportional to magnetization intensity $M$, we fit the data with
formulas
$M(t) = M_{0} + M_{1}\left(\frac{1}{84}e^{- \left( \frac{t}{T_{1}} \right)^{\alpha}} + \frac{3}{44}e^{- \left( \frac{6t}{T_{1}} \right)^{\alpha}} + \frac{75}{364}e^{- \left( \frac{15t}{T_{1}} \right)^{\alpha}} + \frac{1225}{1716}e^{- \left( \frac{28t}{T_{1}} \right)^{\alpha}}\right)$ for
\textsuperscript{139}La and
$M(t)=M_0+M_1 \left(0.028e^{-\left(\frac{t}{T_1} \right)^\alpha }+0.178e^{-\left(\frac{6t}{T_1} \right)^\alpha }+0.794e^{-\left(\frac{15t}{T_1} \right)^\alpha }\right)$
for $^{17}$O, where $t$ is recovery time, and $\alpha$ is stretch
exponent describing a distribution of $T_{1}$. We extracted $T_{1}$
and $\alpha$ from the fitting. Nuclear spin-spin relaxation time
(\(T_{2}\)) measurements were performed with the $\pi/2$-$\tau$-$\pi$-$\tau$
pulse sequence, and we collected the data under different $\tau$ value;
similarly, we can plot $I_s$ versus $\tau$ and fit the
data with the formula $m(\tau) = m_{0} + m_{1}e^{- {\left(\frac{\tau}{T_{2}}\right)}^{\beta}}$ to obtain
$T_{2}$ and $\beta$ for both $^{139}$La and
$^{17}$O.

\subsubsection*{Density Functional Theory (DFT) calculations}

We perform density functional theory (DFT)
calculations~\cite{PhysRev.136.B864, PhysRev.140.A1133}, implemented
in Vienna ab initio simulation package (VASP)
code~\cite{PhysRevB.54.11169} with projector augmented wave (PAW)
method~\cite{PhysRevB.59.1758}. We use Perdew-Burke-Ernzerhof
(PBE)~\cite{PhysRevLett.77.3865} exchange-correlation functional. We
use an energy cutoff of 600 eV and sample the Brillouin zone by using
$\Gamma$-centered \textbf{k}-mesh of 14 $\times$ 14 $\times$ 14 per primitive
cell. The total energy convergence criterion is 10\textsuperscript{-6}
eV. The force convergence criterion is 1 meV/\AA. The pressure
convergence criterion is 0.1 kbar. The empty La-4\emph{f} orbitals are
included in the pseudopotential and we use an on-site Coulomb
interaction U\textsubscript{La-4\emph{f}} = 10 eV to move the
La-\emph{f} orbitals away from the Fermi level. We use
maximally-localized-Wannier-function~\cite{RevModPhys.84.1419},
implemented in Wannier90 ~\cite{MOSTOFI2008685}, to fit the DFT band
structure of LaNiO\textsubscript{2} by the 17-orbital model that
includes one interstitial \emph{s} orbital, five La-\emph{d} orbitals,
five Ni-\emph{d} orbitals, and six O-\emph{p} orbitals. The Wannier
fitting perfectly reproduces the DFT bands structure in an energy
window of about 15 eV around the Fermi level (see Fig. S1 in the
Supplementary Information). Based on the Wannier fitting, we extract
the hopping matrix elements \(t_{\alpha\beta}\) and onsite energy
differences \(\mathrm{\Delta}_{\alpha}\) that will be used in the
cluster model calculation for the superexchange of
LaNiO\textsubscript{2}.

\subsubsection*{Dynamical Mean Field Theory (DMFT) calculations}

Based on the 17-orbital model described above, we perform dynamical mean-field theory (DMFT) calculations~\cite{RevModPhys.68.13, RevModPhys.78.865} to study the mass enhancement of the two Ni-\emph{e\textsubscript{g}} orbitals in LaNiO\textsubscript{2}. The impurity problem is solved using the hybridization-expansion continuous-time quantum Monte Carlo (CTQMC) algorithm~\cite{PhysRevLett.97.076405, RevModPhys.83.349}, as implemented in the code developed by K. Haule~\cite{PhysRevB.75.155113}. At each DMFT iteration, we collect a total of 2 billion Monte Carlo samples to ensure convergence of the impurity Green’s function and self-energy. The calculations are performed at a temperature of 116~K. We employ a Slater-Kanamori interaction on the Ni-\emph{d} orbitals~\cite{PhysRevB.79.115119} and the fully localized limit (FLL) form for the double counting correction~\cite{PhysRevB.49.14211}. The two Ni-\emph{e\textsubscript{g}} orbitals are treated dynamically within DMFT, while the filled Ni-\emph{t\textsubscript{2g}} shell is handled at the static Hartree-Fock level. Because the system is metallic, the quasiparticle weight \(Z\), defined via the real part of the retarded self-energy, can be used to estimate the mass enhancement:
\begin{equation}
    \frac{m^{*}}{m} \approx \frac{1}{Z} = 1 - \left. \frac{1}{\hbar} \frac{\partial \operatorname{Re} \Sigma(\omega)}{\partial\omega} \right|_{\omega = 0}
\end{equation}
Here, \(\omega = 0\) denotes the chemical potential. Since CTQMC yields the self-energy on the Matsubara axis, we infer the low-frequency behavior under the assumption of Fermi-liquid behavior, which allows us to estimate \(Z\) from the imaginary part of the self-energy at low Matsubara frequencies~\cite{PhysRevB.91.195149}:
\begin{equation}
    \frac{m^{*}}{m} \approx \frac{1}{Z} \approx 1 - \left. \frac{1}{\hbar} \frac{\partial \operatorname{Im} \Sigma(i\omega_{n})}{\partial\omega_{n}} \right|_{\omega_{n} \to 0}
\end{equation}
In practice, we fit a fourth-order polynomial to the first six Matsubara-frequency points of \(\textrm{Im}\Sigma(i\omega_{n})\) and compute the derivative from this fit. For \(U = 5\)~eV and \(J_{H} = 0.2U\), we obtain a mass enhancement of 5.4 for the Ni-\(d_{x^{2} - y^{2}}\) orbital, in agreement with previous GW+EDMFT results~\cite{PhysRevX.10.041047}. We therefore adopt \(U = 5\)~eV and \(J_{H} = 0.2U\) as representative interaction parameters. The corresponding Matsubara self-energy is shown in Fig.~S6 of the Supplementary Information. We have verified that the main conclusions remain robust across a range of \(U\) and \(J_{H}\) values.

\subsubsection*{Modeling the superexchange of LaNiO$_2$}

We extend the classical three-orbital $pd$ model to a $pds$ model, schematically illustrated in Fig.~\ref{fig4}(b). This model comprises nine orbitals: two Ni-$d_{x^2-y^2}$ orbitals, two Ni-$d_{3z^2-r^2}$ orbitals, one O-$p_x$ orbital, and four interstitial $s$ orbitals. The model is defined by the Hamiltonian:
\begin{equation}
	\label{eq1} \hat{H} = \hat{H}_0 + \hat{H}_{\rm{int}},
\end{equation}
where $\hat{H}_0$ is the tight-binding part:
\begin{eqnarray}
	\label{eq2} \hat{H}_0 &=& \sum_{\sigma} \Delta_{p_x} \hat{c}^{\dagger}_{p_x \sigma}\hat{c}_{p_x \sigma} +  \sum_{\substack{\sigma \\ i=\{1,2\}}} \Delta_{d_z} \hat{c}^{\dagger}_{i,d_z \sigma}\hat{c}_{i, d_z\sigma} + \sum_{\substack{\sigma \\ j=\{\textrm{1t, 1b, 2t, 2b}\}}} \Delta_s \hat{c}^{\dagger}_{j,s \sigma}\hat{c}_{j, s\sigma} \\ \nonumber
	&&+ \sum_{\substack{\sigma \\ i=\{1,2\}}} t_{p_x d_x} \left( (-1)^i \hat{c}^{\dagger}_{i, d_x \sigma}\hat{c}_{p_x \sigma} + \textrm{h.c.} \right) + \sum_{\substack{\sigma \\ i=\{1,2\}}} t_{p_x d_z} \left( (-1)^i \hat{c}^{\dagger}_{i, d_z \sigma}\hat{c}_{p_x \sigma} + \textrm{h.c.} \right) \\ \nonumber
	&&+ \sum_{\substack{\sigma \\ j=\{\textrm{1t,1b}\}}} t_{s d_z} \left( \hat{c}^{\dagger}_{j, s \sigma}\hat{c}_{1, d_z \sigma} + \textrm{h.c.} \right) + \sum_{\substack{\sigma \\ j=\{\textrm{2t,2b}\}}} t_{s d_z} \left( \hat{c}^{\dagger}_{j, s \sigma}\hat{c}_{2, d_z \sigma} + \textrm{h.c.} \right) \\ \nonumber
	&&+ \sum_{\substack{\sigma \\ j=\{\textrm{1t,1b}\}}} t_{s p_x} \left( \hat{c}^{\dagger}_{j, s \sigma}\hat{c}_{p_x \sigma} + \textrm{h.c.} \right) + \sum_{\substack{\sigma \\ j=\{\textrm{2t,2b}\}}} \left(-t_{s p_x}\right) \left( \hat{c}^{\dagger}_{j, s \sigma}\hat{c}_{p_x \sigma} + \textrm{h.c.} \right)
\end{eqnarray}
Here, ``1'' and ``2'' label the two Ni sites, and ``t'' and ``b'' refer to ``top'' and ``bottom''. $\hat{H}_{\rm{int}}$ is a Slater–Kanamori interaction on the $d$ orbitals of both Ni sites:
\begin{eqnarray}
	\label{eqSr} \hat{H}_{\rm{int}} &=& \sum_{\substack{i=\{1,2\}\\ \nu={d_x, d_z}}} U \hat{n}_{i, \nu\uparrow}\hat{n}_{i, \nu \downarrow} + \sum_{\substack{\sigma\sigma' \\ i=\{1,2\}}} \big(U' - J_H \delta_{\sigma \sigma'}\big) \hat{n}_{i, d_x\sigma} \hat{n}_{i, d_z\sigma'} \\ \nonumber
	&&+ \sum_{i=\{1,2\}} (-J_H)\left( \hat{c}^{\dagger}_{i, d_z \downarrow}\hat{c}^{\dagger}_{i, d_x \uparrow}\hat{c}_{i, d_x \downarrow}\hat{c}_{i, d_z \uparrow} + \textrm{h.c.} \right) \\ \nonumber
	&&+ \sum_{i=\{1,2\}} (-J')\left( \hat{c}^{\dagger}_{i, d_z \downarrow}\hat{c}^{\dagger}_{i, d_z \uparrow}\hat{c}_{i, d_x \downarrow}\hat{c}_{i, d_x \uparrow} + \textrm{h.c.} \right)
\end{eqnarray}
Here, $d_x$ ($d_z$) denotes Ni-$d_{x^2 - y^2}$ (Ni-$d_{3z^2 - r^2}$) orbital. The operator $\hat{c}^{\dagger}_{i,\nu \sigma}$ creates an electron at site $i$ in orbital $\nu$ with spin $\sigma$. ``h.c.'' indicates the Hermitian conjugate. The model parameters are defined as follows: in $\hat{H}_0$, $\Delta_{\alpha}$ is the onsite energy of orbital $\alpha$ relative to Ni-$d_{x^2 - y^2}$ (thus $\Delta_{d_x} = 0$ and is not shown); $t_{\alpha \beta}$ denotes the nearest-neighbor hopping between orbitals $\alpha$ and $\beta$. In $\hat{H}_{\rm{int}}$, $U$ is the intra-orbital Coulomb interaction, $U'$ is the inter-orbital interaction, $J_H$ is the Hund’s coupling, and $J'$ is the pair-hopping term. To preserve spin SU(2) symmetry, we set $U' = U - 2J_H$ and $J' = J_H$. The values of $\Delta_{\alpha}$ and $t_{\alpha\beta}$ are listed in Table~\ref{tab:1}.

\begin{table*}[t!]
  \caption{Model parameters. $d_x$ ($d_z$) denotes Ni-$d_{x^2 - y^2}$ (Ni-$d_{3z^2 - r^2}$) orbital. \(\mathrm{\Delta}_{\alpha}\) is the onsite energy of orbital \(\alpha\) relative to Ni-$d_{x^{2} - y^{2}}$; \(t_{\alpha\beta}\) is the hopping between orbitals \(\alpha\) and \(\beta\). Units are in eV.}
	\begin{ruledtabular}
		\begin{tabular}{cccc}
			$\Delta_{p_x}$ & $\Delta_s$ & $\Delta_{d_z}$ \\ \colrule
			-3.77 & 2.59 & -0.04 \\ \colrule
			$t_{p_x d_x}$ & $t_{p_x d_z}$ & $t_{s p_x}$ & $t_{s d_z}$ \\ \colrule
			1.27 & 0.66 & 0.57 & 1.08
		\end{tabular}
	\end{ruledtabular}
	\label{tab:1}
\end{table*}

In the $pds$ model with $N_{\rm{orb}} = 9$, we set the total electron number to $N_e = 8$. This reflects the nominal $d^9$ configuration of Ni in LaNiO$_2$, corresponding to three electrons in the two $e_g$ orbitals on each Ni site. The O-$p_x$ orbital is formally filled with two electrons, and the interstitial $s$ orbitals are nominally empty. Since the system effectively contains two coupled spin-$\frac{1}{2}$ moments, we define ``ferromagnetic'' and ``antiferromagnetic'' configurations as follows.

For the ``ferromagnetic'' configuration, we fix the total spin projection $S_z = 1$, resulting in $N^{\uparrow}_e = 5$ and $N^{\downarrow}_e = 3$. The corresponding many-body Hilbert space has dimension $C_{N_{\rm{orb}}}^{N^{\uparrow}_e} \times C_{N_{\rm{orb}}}^{N^{\downarrow}_e} = 10{,}584$. For the “antiferromagnetic” configuration, we set $S_z = 0$, giving $N^{\uparrow}_e = N^{\downarrow}_e = 4$, and a Hilbert space dimension of $15{,}876$.

We construct the Hamiltonian in Eq.~(\ref{eq1}) within both configurations and perform exact diagonalization to obtain the ground-state energies $E_{\rm{FM}}$ and $E_{\rm{AFM}}$. We then map the $pds$ model onto a spin-$\frac{1}{2}$ Heisenberg model, $H = J \mathbf{S}_1 \cdot \mathbf{S}_2$, where the energy difference between the spin-singlet and spin-triplet states is $-J$. Therefore, the superexchange coupling is determined as $J = |E_{\rm{FM}} - E_{\rm{AFM}}|$, with $E_{\rm{AFM}} < E_{\rm{FM}}$.

\section*{Acknowledgements}

We acknowledge useful discussions with Mi Jiang, Mona Berciu and George A. Sawatzky. This work is supported by the National Key R\&D Program of the MOST of China (Grant No. 2022YFA1602601), the National Natural Science Foundation of China (Grants No. 12034004, 12161160316, 12325403, 12488201,12204450), the Chinese Academy of Sciences under contract No. JZHKYPT-2021-08, the CAS Project for Young Scientists in Basic Research (Grant No. 2022YSBR-048), the Innovation Program for Quantum Science and Technology (Grant No. 2021ZD0302800), and the Fundamental Research Funds for the Central Universities (No. WK9990000110). H.H.C. was financially supported by the National Natural Science Foundation of China under project number 12374064 and 12434002, Science and Technology Commission of Shanghai Municipality under grant number 23ZR1445400 and a grant from the New York University Research Catalyst Prize under project number RB627. C.L.X. was supported by the National Natural Science Foundation of China under project number 12404082. NYU High-Performance-Computing (HPC) provides computational resources.

\section*{Author contributions}
X.H.C., T.W. and H.H.C. conceived the project. Y.B.Z.,  D.Z. and Y.W. performed NMR experiments. Y.B.Z. grew the samples with the help of D.Z.. B.Y.Z., C.L.X. and H.H.C. performed theoretical calculations.Y.B.Z., D.Z. and T.W. analysed the NMR data. Y.B.Z., D.Z., H.H.C. and T.W. wrote the manuscript with input from 
all the authors. All authors discussed the results and commented on the manuscript.

\section*{Competing interests}
The authors declare no competing interests.

\newpage

\begin{figure}[t]
	\includegraphics[angle=0,width=0.95\textwidth]{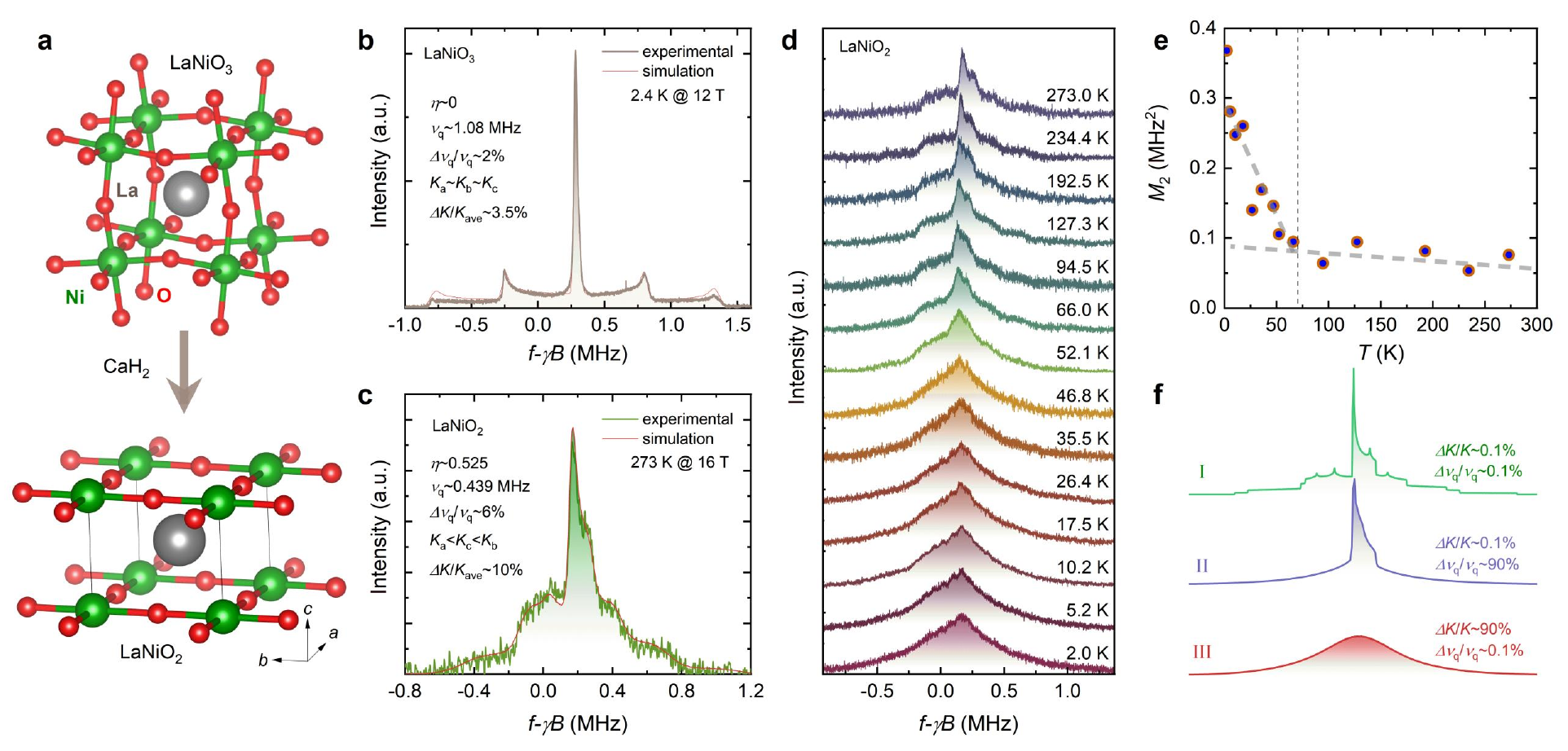}
	\caption{\label{fig1} \textbf{Crystal structure and
			temperature-dependent $^{17}$O NMR spectra in
			LaNiO$_{2}$.}
		(a) Sketch of the structural change from LaNiO$_3$ to LaNiO$_{2}$ by topotactic reduction with metal hydride CaH$_2$. (b-c) NMR spectra
		of LaNiO$_3$ and LaNiO$_2$. The solid red curve represents the simulation result. The asymmetry factor
		$\eta =\left| \frac{V_{cc}-V_{aa}}{V_{bb}}\right|$   characterizes
		the asymmetry of the electric field gradient, where
		$V_{\alpha\alpha} = \frac{\partial^{2}V}{\partial\alpha^{2}}$
		($\alpha = a$, $b$, and $c$) represents the second
		derivative of the electric potential at the $^{17}$O
		nuclei along the three crystallographic directions. $K_a$, $K_b$ and $K_c$ describe
		the Knight shifts along the three crystallographic directions (for details, see Sec. S3 in the Supplemental Information). (d) $^{17}$O full spectra of LaNiO$_2$ at temperatures from 2.0 K to 273.0 K at a fixed field of $B\sim16$ T. (e) Temperature-dependent second central moment of the $^{17}$O spectra of LaNiO$_2$. (f) Simulation of the broadening effect in $^{17}$O spectra. 
		}
\end{figure}

\begin{figure}[t]
	\includegraphics[angle=0,width=0.95\textwidth]{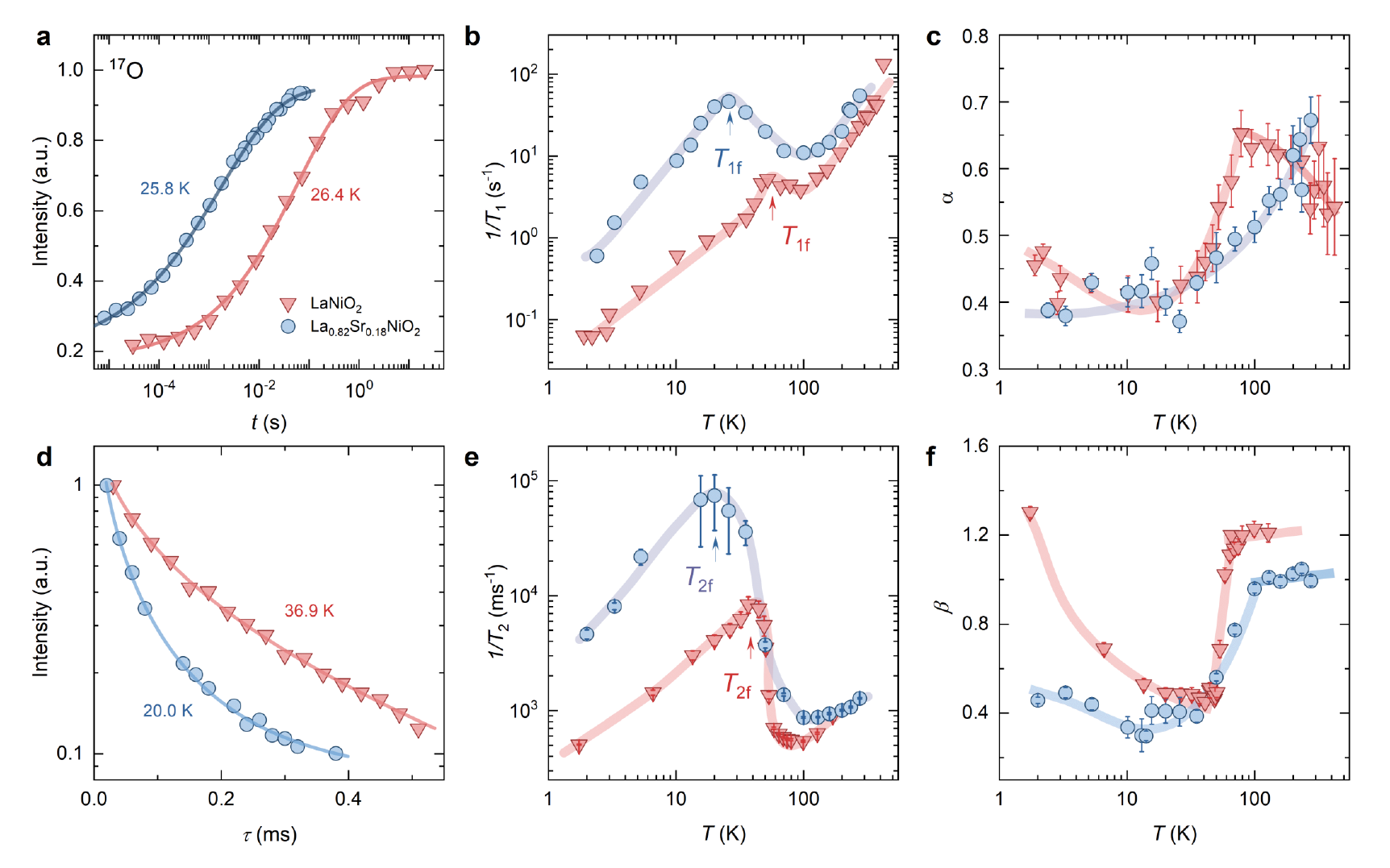}
	\caption{\label{fig2} \textbf{NMR relaxation evidence for glassy spin dynamics in LaNiO$_2$ and La$_{0.82}$Sr$_{0.18}$NiO$_2$.}
		(a) Time dependence of  $^{17}$O
		nuclear magnetization \(M(t) = M_{z}(t)/M_{z}(\infty)\) after a
		comb of $\pi/2$ saturation pulses in LaNiO\textsubscript{2} and
		La\textsubscript{0.82}Sr\textsubscript{0.18}NiO\textsubscript{2}. By fitting the relaxation curves via the stretched exponential form $M(t)=M_0+M_1\left( 0.028e^{-\left(\frac{t}{T_1} \right)^\alpha }+0.178e^{-\left(\frac{6t}{T_1} \right)^\alpha }+0.794e^{-\left(\frac{15t}{T_1} \right)^\alpha } \right)$, we extracted both the spin-lattice relaxation time
		$T_1$ and the stretching exponent $\alpha$. (b) The temperature-dependent spin-lattice relaxation rate \(1/T_{1}\) for
		\textsuperscript{17}O in LaNiO\textsubscript{2} and
		La\textsubscript{0.82}Sr\textsubscript{0.18}NiO\textsubscript{2} (c) The temperature-dependent stretching factor \(\alpha\) in
		LaNiO\textsubscript{2} and
		La\textsubscript{0.82}Sr\textsubscript{0.18}NiO\textsubscript{2}. (d) The spin-echo decay curves of \textsuperscript{17}O in LaNiO\textsubscript{2} and
		La\textsubscript{0.82}Sr\textsubscript{0.18}NiO\textsubscript{2}. (e) The temperature-dependent spin-spin relaxation rate
		\(1/T_{2}\) for \textsuperscript{17}O in LaNiO\textsubscript{2} and
		La\textsubscript{0.82}Sr\textsubscript{0.18}NiO\textsubscript{2}. (f) The temperature-dependent factor \(\beta\) in
		LaNiO\textsubscript{2} and
		La\textsubscript{0.82}Sr\textsubscript{0.18}NiO\textsubscript{2}
		}
\end{figure}

\begin{figure}[t]
	\includegraphics[angle=0,width=0.95\textwidth]{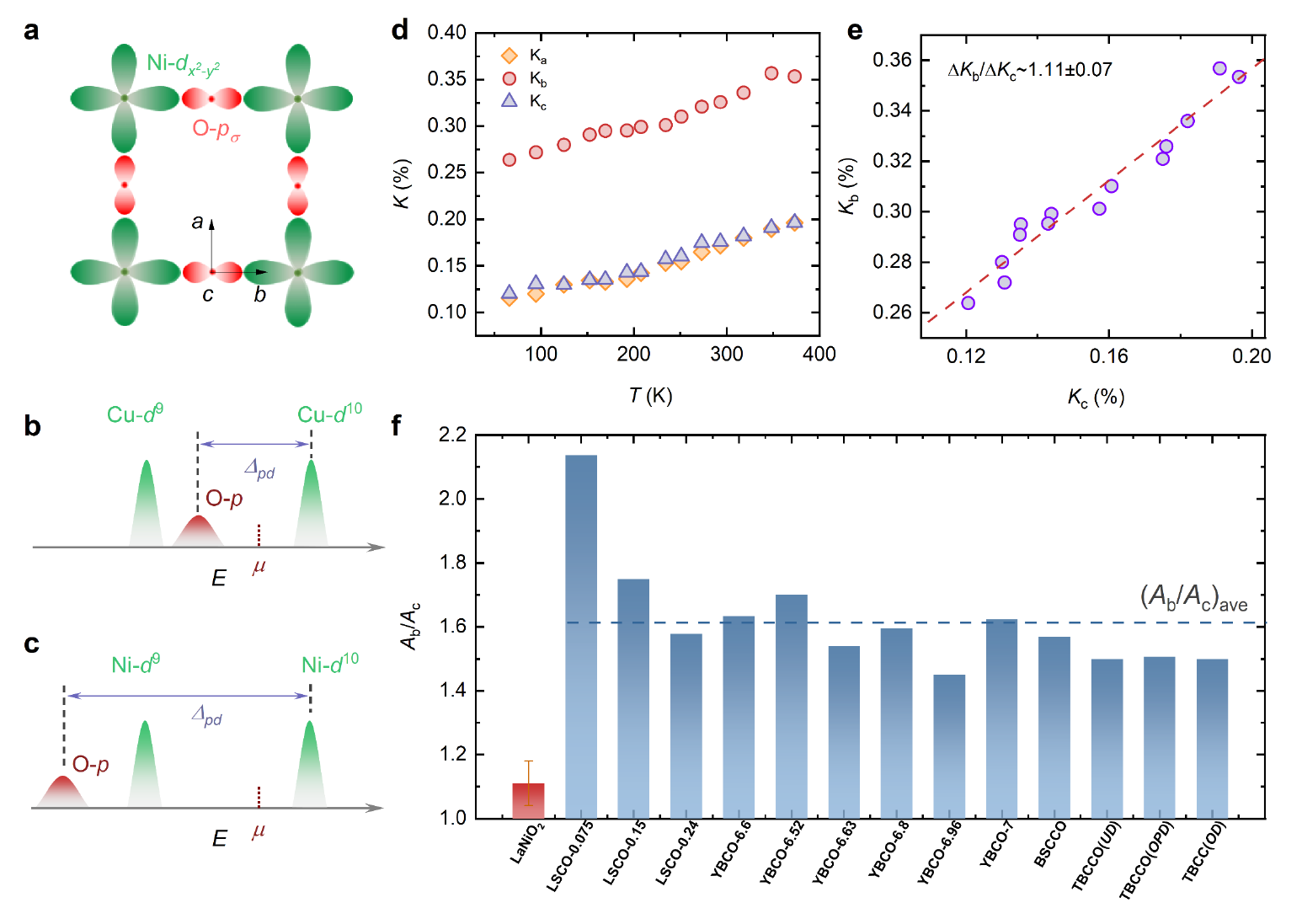}
	\caption{\label{fig3}
		\textbf{}\textbf{Knight shift analysis of superexchange interactions.} (a) Sketch of
		Ni-$d$ orbitals and O-$p$ orbitals in NiO\textsubscript{2}
		plane. (b)-(c) Schematic of the Zaanen-Sawatzky-Allen electronic structure
		comparison between the infinite-layer nickelates and cuprates, showing
		the O-$p$, Ni-$d$ and Cu-$d$ states. This illustrates the increase
		in nickelate charge-transfer energy $\Delta_{pd}$ relative to the
		Coubomb repulsion $U$. (d) Temperature-dependent Knight shift
		\textsuperscript{17}\emph{K} along \emph{a}, \emph{b} and \emph{c}
		directions. (c) \textsuperscript{17}\emph{K\textsubscript{b}}
		versus \textsuperscript{17}\emph{K\textsubscript{c}}, which are
		extracted from the data in panel (d). The dashed line is a
		linear fit whose slope is \(1.11 \pm 0.07\). (f) Ratios
		of hyperfine coupling constants along \emph{b} and \emph{c} directions
		for LaNiO\textsubscript{2} in this work, as well as for a variety of
		cuprates based on literature data~\cite{takigawa1989oxygen,zheng1993oxygen,yoshinari1990nmr,horvatic1993nuclear,martindale1998temperature,crocker2011nmr,gerashenko1999cu}. The blue dashed line
		shows the mean value of the \(A_{b}/A_{c}\) for cuprates.}
\end{figure}

\begin{figure}[t]
	\includegraphics[angle=0,width=0.95\textwidth]{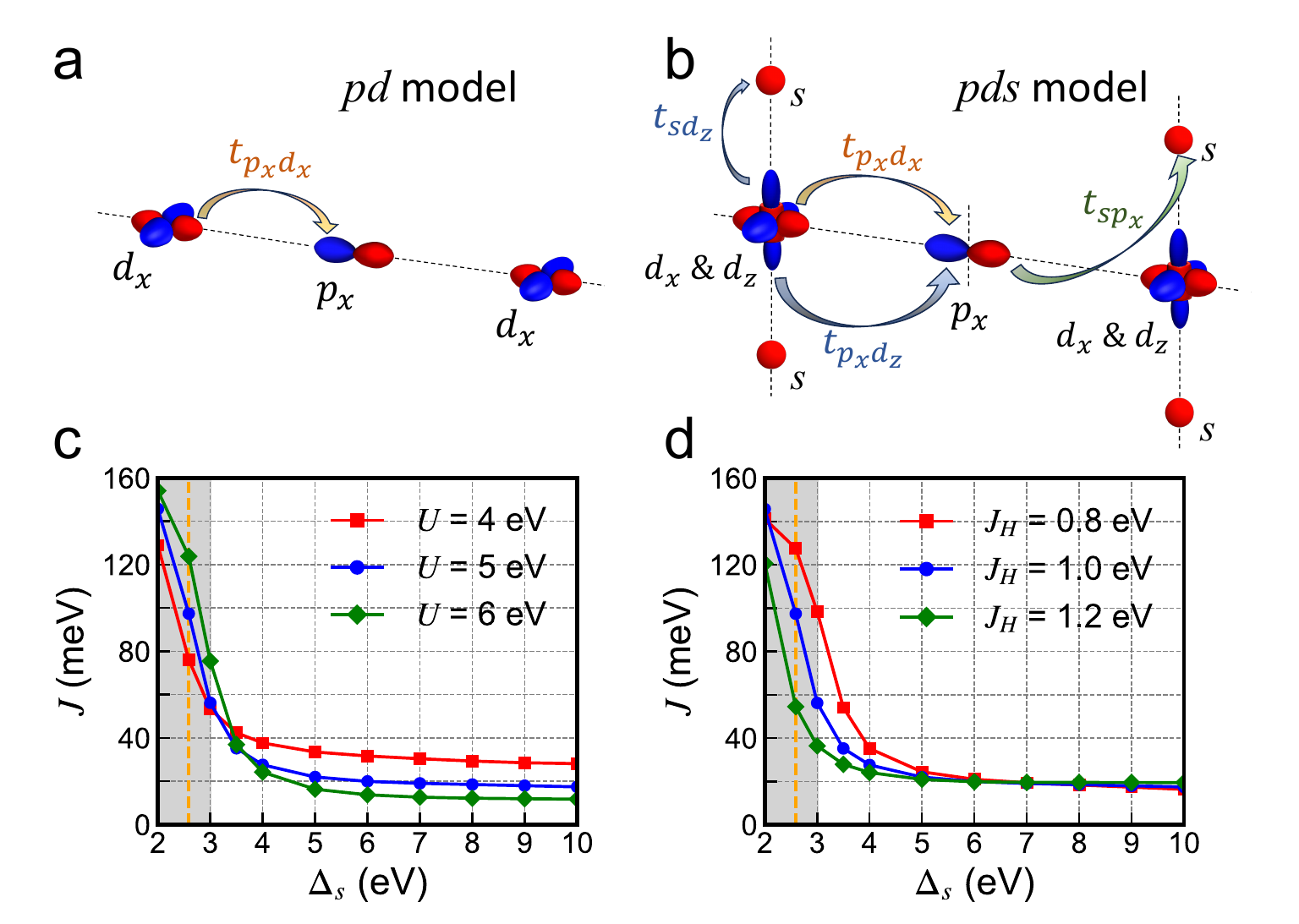}
	\caption{\label{fig4} \textbf{Modeling the superexchange in
            LaNiO\(_2\).}  (a) The \( pd \) model, consisting of two
          \( d_{x^2 - y^2} \) orbitals and a central \( p_x \)
          orbital, capturing the conventional superexchange pathway
          via the $pd$ $\sigma$-bond.  (b) The extended \( pds \)
          model, comprising two correlated Ni sites. Each site hosts a
          Ni-$d_{x^2 - y^2}$ and a Ni-$d_{3z^2 - r^2}$ orbital, and is
          flanked by two interstitial $s$ orbitals. A single O-$p_x$
          orbital is located between the two Ni sites. In both (a) and
          (b), \( d_x \) and \( d_z \) denote the Ni-$d_{x^2 -y^2}$
          and Ni-$d_{3z^2 - r^2}$ orbitals, respectively. All
          symmetry-allowed nearest-neighbor hoppings \(
          t_{\alpha\beta} \) between orbitals \( \alpha \) and \(
          \beta \) are explicitly shown.  (c) Superexchange coupling
          \( J \) as a function of the interstitial orbital energy \(
          \Delta_s \), calculated for several values of the Hubbard
          interaction \( U \), with Hund's coupling fixed at \( J_H =
          0.2U \).  (d) Superexchange \( J \) as a function of \(
          \Delta_s \), calculated for varying \( J_H \) values at
          fixed \( U = 5\,\mathrm{eV} \).}
\end{figure}

\clearpage
\newpage
\bibliography{ref}

\end{document}